# Traffic flow clustering framework using drone video trajectories to automatically identify surrogate safety measures

**Shengxuan Ding**
Department of Civil, Environmental & Construction Engineering
University of Central Florida, Orlando, FL, 32816, USA
Email: Shengxuan.Ding@ucf.edu

**Mohamed Abdel-Aty, PhD**
Department of Civil, Environmental & Construction Engineering
University of Central Florida, Orlando, FL, 32816, USA
Email: m.aty@ucf.edu

**Ou Zheng, PhD (Corresponding author)**
Department of Civil, Environmental & Construction Engineering
University of Central Florida, Orlando, FL, 32816, USA
Email: Ou.Zheng@ucf.edu

**Zijin Wang**
Department of Civil, Environmental & Construction Engineering
University of Central Florida, Orlando, FL, 32816, USA
Email: Zijin.Wang@ucf.edu

**Dongdong Wang, PhD**
Department of Civil, Environmental & Construction Engineering University of Central Florida, Orlando, FL, 32816, USA
Email: Dongdong.Wang@ucf.edu

Word Count: 5,628 words + 3 table (750 words per table) = 6,378 words

*Submitted [07/27/2023]*

**ABSTRACT**
The utilization of traffic conflict indicators is crucial for assessing traffic safety, especially when the crash data is unavailable. To identify traffic conflicts based on traffic flow characteristics across various traffic states, we propose a framework that utilizes unsupervised learning to automatically establish surrogate safety measures (SSM) thresholds. Different traffic states and corresponding transitions are identified with the three-phase traffic theory using high-resolution trajectory data.  Meanwhile, the SSMs are mapped to the corresponding traffic states from the perspectives of time, space, and deceleration. Three models, including k-means, GMM, and Mclust, are investigated and compared to optimize the identification of traffic conflicts. It is observed that Mclust outperforms the others based on the evaluation metrics. According to the results, there is a variation in the distribution of traffic conflicts among different traffic states, wide moving jam (phase J) has the highest conflict risk, followed by synchronous flow (phase S), and free flow (phase F). Meanwhile, the thresholds of traffic conflicts cannot be fully represented by the same value through different traffic states. It reveals that the heterogeneity of thresholds is exhibited across traffic state transitions, which justifies the necessity of dynamic thresholds for traffic conflict analysis.

**INTRODUCTION**

Over the last two decades, the severity of crashes has been extensively studied to understand the frequency and causes. Previous studies have attempted to assess the real-time likelihood of risk occurrence by identifying factors that occurred before the crash *(1)*. However, using historical crash data for safety evaluation has issues with risk mitigation actions due to incorrect reasoning, subjectivism, and inaccurate data *(2)*. Moreover, factors resulting in a crash like driver behavior may not be modeled using the prediction algorithms in the literature *(3)*. On the other hand, SSMs and microscopic traffic data have been proven to be appealing and widely used for analyzing traffic safety performance *(4)*. Hence, the goal of this study is to identify conflicts and link them to traffic flow characteristics using empirical trajectory data.

The first objective of this paper is to explore the mechanism of traffic conflicts from the perspective of macroscopic traffic states. It has been demonstrated that dynamic traffic states with spatiotemporal characteristics are effective as crash precursors for conflict detection. Traffic flow characteristics significantly impact traffic safety performance *(5)*, which can be divided into various categories to identify the factors of traffic flow and how they relate to conflicts. Levels of service (LOS), three-phase theory, and fundamental diagram are three typical methods to represent different traffic flow levels. Many studies considered heterogeneity at different macroscopic traffic states in combination with classified real-time safety performance *(6)*. Because macroscopic traffic flow has a significant impact on traffic conflicts, the transition between traffic states and the relationship of traffic parameters are critical to the causality of traffic conflicts. Most conflict forecasts rely on analyzing the micro-traffic flow features rather than comprehending the mechanism from the perspective of macro-traffic flow characteristics. Simultaneously, the relationship between traffic state and conflicts must be thoroughly investigated from both macro and micro levels. This can be accomplished using trajectory data rather than aggregate-level data by analyzing vehicle interactions and conflict risks *(7)*.

The second major objective of this paper is to identify traffic conflicts and determine SSM indicator thresholds using empirical trajectory data. The dataset employed in this study is a portion of the CitySim dataset *(8)*, which comprises a higher frequency of conflicts and diverse traffic states in freeway segments. The dataset captures vehicle trajectories with the coordinates of four bounding box points and the center point of vehicles. Utilizing these points to compute the precise value of SSMs holds immense significance, as it circumvents the errors caused by vehicle size when solely relying on the center point of each vehicle. In contrast to CCTV or roadside camera footage, this dataset provides complete coverage of the freeway segment and incorporates an aerial perspective to rectify angle-related errors. Additionally, the raw trajectory data contains comprehensive details about individual vehicles at each frame, including their speed, direction, and lane identification. The high quality of this dataset provides a solid foundation for the analysis of traffic conflicts. However, few studies have concentrated on conflict mechanisms and inherent heterogeneity in traffic flow characteristics with high-resolution trajectory data. Most studies predicted conflicts and risk utilizing microscopic kinematics features between adjacent vehicles rather than macroscopic traffic flow features *(9)*. Dynamic thresholds are necessary for traffic conflict analysis due to the heterogeneity of thresholds across traffic state transitions. The selection of different thresholds of various scenarios can help us better understand the correlation between traffic conflicts and traffic flow parameters. Along these lines, the use of dynamic thresholds for conflict analysis is an interesting avenue to explore with the algorithm development for automated vehicles (AVs). This framework can provide a comprehensive assessment of the severity of conflicts for AVs by considering the proximity to collision and traffic states.

The remaining research is listed below. Section 2 included a literature review. The methodology is covered in Section 3. Section 4 contains the results and discussions. Section 5 discusses the findings of the study as well as future research.

**LITERATURE REVIEW**

The performance of traffic safety is heavily influenced by traffic flow, several studies discuss the correlation between crash and traffic flow. A link between traffic characteristics and daytime freeway crashes is established to confirm the importance of flow variation in traffic safety *(10)*. High-resolution trajectory data is applied to evaluate heterogeneous crash mechanisms under different traffic states *(11)*. However, crashes may not occur in many conditions, where safety evaluation is dependent on traffic flow characteristics and traffic conflicts. Probabilistic neural network (PNN) models for identifying and predicting rear-end collisions at congested flow and free flow based on loop detector data were estimated separately *(12)*. A logit model with random parameters and heterogeneity in means and variances was used to investigate the relationship between conflicts and traffic flow characteristics *(13)*. In previous research on traffic flow and states, a three-phase traffic flow theory is developed based on expressway data *(14)*. The three-phase traffic flow theory divided traffic states into three categories: free flow(F), synchronous flow(S), and wide moving jam(J). When in a free flow state, traffic flow is at a low density and high speed without disturbance of other vehicles. Furthermore, high flow and speed distinguish synchronous flow. Compared with free flow, the average speed is slower, and the density is higher. While in a wide moving jam flow, the flow and speed tend to be zero, and the density reaches a maximum. Transition processes exist for free flow and synchronous flow (F—S), free flow and wide moving jam (F—J), and synchronous flow and wide moving jam (S—J). Transitions between these three phases can all be first-order transitions. Among them, the transition from free flow to wide moving jam requires two steps. First, free flow transforms into synchronous flow, which then generates wide moving jams. Based on the three-phase theory, traffic states and variables can be chosen to assess the relationship between traffic flow and safety performance *(15)*. Traffic safety analysis with three-phase traffic flow theory is conducted with aggregated traffic flow data *(16)*.

    To comprehensively examine traffic safety concerns that involve drivers, vehicles, and roads, safety surrogate measures (SSMs) have been widely used to measure different dimensions of conflicts *(17)*. The benefits and drawbacks of various SSMs are summarized *(18)*. The performance of SSMs by six indexes to calibrate threshold with naturalistic driving data is assessed *(19)*. Fuzzy Surrogate Safety Metrics are presented to distinguish between safe and unsafe situations for rear-end collision *(20)*. Modified SSMs were used to capture the probability and severity of collisions based on simulation *(21)*. Traffic safety is assessed at signalized intersections by simulator validity from perspectives of traffic and safety parameters *(22)*. A series of machine learning algorithms and statistical learning techniques are generally applied to determine factors of conflict and predict the occurrence of conflict. Different network models such as CNN *(23)*, LSTM *(24)*, and DNN *(25)* were proposed to detect and predict traffic conflicts from traffic variables and SSM. Based on statistical methods, conditional logistic regression *(26)*, stratified sampling *(27)*, and multiple logistic regression models *(28)* were used to estimate conflict risk. In addition, the Peak Over Threshold (POT) approach *(29)* and Multivariate Extreme Value models *(30)* were also widely used to identify conflict frequency.

    Identifying traffic conflicts and determining thresholds are key for assessing safety performance. Typically, previous studies define thresholds with one value, disregarding their suitability for their studies. Even within the same context, multiple thresholds were proposed for analyzing conflicts. For example, the range of TTC thresholds varied widely from 0.5 s to 6.0s at signalized intersections for rear-end *(31)*. It is observed the same problem with PET thresholds as well *(32)*. Given the wide variation in the prescribed surrogate thresholds, some researchers have estimated the thresholds empirically. There are several major approaches for measuring conflict thresholds as shown in **Table 1**. While some studies determine the threshold of SSM based on real crash data, the selection of thresholds of various scenarios can be significantly different. This work addresses this research gap and uses high-resolution trajectory data to analyze three-phase traffic states by different traffic flow characteristics. The main objective of this work was to propose clustering methods, imbalanced data processing, and unsupervised learning evaluation on conflict identification and threshold selection for future research. Thus, these contributions can be applied in different types of locations at various traffic states. This can help us better understand the correlation between traffic conflicts and traffic flow parameters, which may be applied to investigate the differences and associations between microscopic conflict and macroscopic traffic flow.

**TABLE 1 Methods to measure conflict thresholds**

| Method | Description | Advantage | Disadvantage |
|---|---|---|---|
| Correlational approach | (1) Cumulative density function (CDF) <br> (2) Extreme Value models and the observed crashes <br> (3) ROC (Receiver Operating Characteristics) curves | (1) Increase the detection of valid events. <br> (2) Reduce the number of false positives. | (1) Ignore a significant portion of the existing events. <br> (2) An external parameter may influence parameter estimation. <br> (3) Weaken safety-relevant meaning in reality |
| Distribution-based approach | (1) Bimodal histogram method <br> (2) Percentile method <br> (3) Deviation method | (1) Using microscopic traffic data to predict collisions using binary classification | (1) Spatial and temporal transferability |
| Classification methods | (1) Linear discriminant analysis <br> (2) Discrete choice modelling <br> (3) Machine learning | (1) Detect the occurrence of conflicts between surrounding vehicles accurately. <br> (2) Consider the dependencies between the vehicles and networks | (1) Need to develop a single value regarding safety for each approach. <br> (2) Investigate changes of thresholds between different sites |
| Time Series Analysis | (1) Spectral clustering <br> (2) Discrete Fourier Transform <br> (3) Permutation entropy of kinematic indicators | (1) Inability to distinguish between crashes and near-crashes <br> (2) Incorporate additional variables | (1) Determine driving volatility by quantifying the complexity of data <br> (2) Remove fixed-zone detection constraints and multiple computations at each time step |
| Extreme value estimation | (1) Mean residual life plot <br> (2) Threshold stability plot | (1) Explain the heterogeneity in thresholds <br> (2) Estimation does not require crash data | (1) Requires high-quality traffic conflict and crash data over a longer period <br> (2) Sensitivity and specificity need to be improved |

**DATA PREPARATION**

The researchers of this study opted to employ the "Citysim Dataset" *(8)*, an open-source dataset known for its remarkably high resolution of 4K (4096 × 2160) at 30 frames per second, captured from drone videos. **Figure 1** illustrates a schematic diagram and an aerial view of the research area. The study area covers 680 m in length and consists of six lanes. During peak hours, a total of 35 minutes of data is available for the entire sample segment, divided into two periods: (1) 5:20 p.m. to 5:35 p.m., and (2) 5:48 p.m. to 6:07 p.m. To begin, vehicles are separated by lanes because traffic conditions differ between lanes.

*Ding, Abdel-Aty, Zheng, Wang, and Wang*

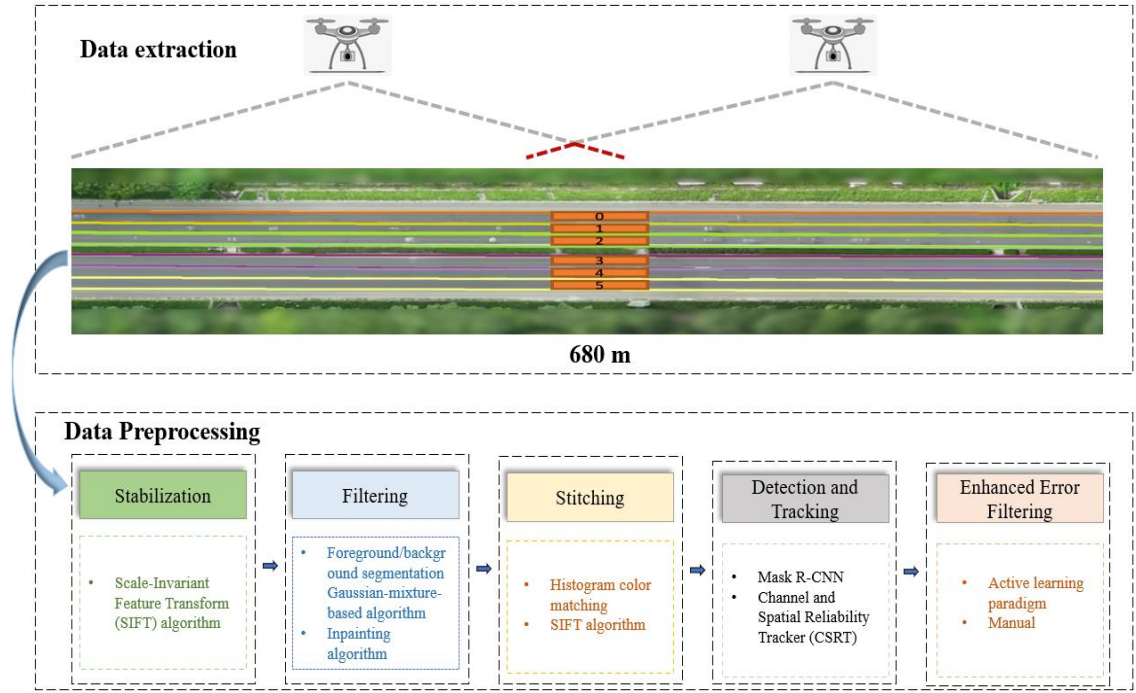

**Figure 1 Aerial view of freeway segment and flow chart of data preprocessing**

    Following that, we filter out all lane-changing and cut-in behavior to focus on rear-end conflicts. To reduce noise, every 30 frames (1 second) are aggregated to calculate the precise value of traffic characteristics and moving average method is also applied to smooth the data. Vehicles in each sub-segment are paired to ensure that the trajectory remains continuous. Furthermore, the first and last vehicles in each video are removed. Following the preprocessing of data, many traffic flow characteristics are calculated to identify traffic states. We select space speed as one of the indexes to evaluate macro traffic states, which is calculated by the average speed of all vehicles in the road segment. Density is defined as the number of vehicles divided by the length of the road segment. What's more, the flow rate is formulated by the average time headway: $q(flow\ rate) = 1/\bar{h}(time\ headway)$. Vehicular trajectories to westbound with instantaneous speeds are shown in **Figure 2**.

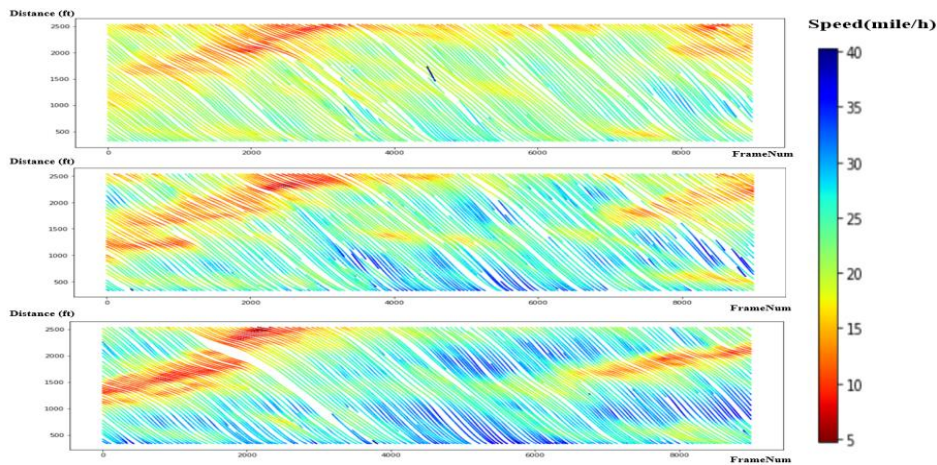

**Figure 2 Vehicular trajectories with instantaneous speeds**

**METHODS**

The methodology of this paper is described in the following subsections: identification of traffic states, calculation of SSM, clustering methods, and evaluation. Initially, vehicle trajectory data is analyzed for freeway segments, and traffic flow variables such as flow rate, density, and average speed are calculated to classify traffic states according to the three-phase theory framework. Subsequently, the SSMs are computed for further study. Finally, unsupervised learning models including k-means, GMM, and Mclust are compared to automatically establish SSM thresholds. **Figure 3** illustrates the proposed methods.

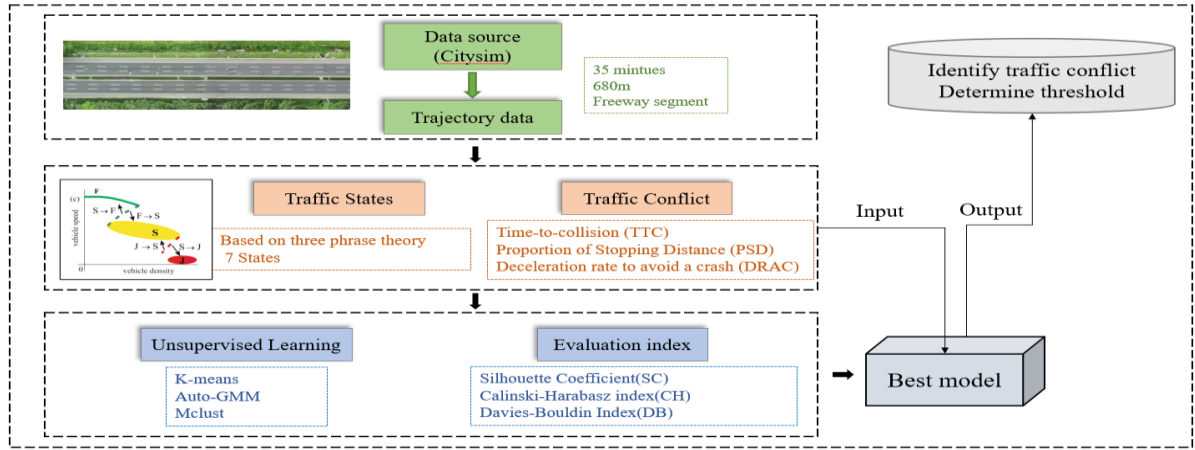

**Figure 3 Diagram of the methodology**

**Identification of traffic state**

The objective of this work is to connect the divide between conflict and traffic flow features and to broaden the conventional conflict risk assessment to encompass the traffic flow condition. **Figure 4** provides a visual representation of the criteria for traffic state identification. The traffic phase in a wide-moving jam can be determined by analyzing a time series plot that shows the effects of speed, time headway, and traffic flow interruptions. This analysis involves using several criteria, such as the average speed, maximum time headway, correlation coefficient between density and flow rate, and the number of vehicles in the phase.

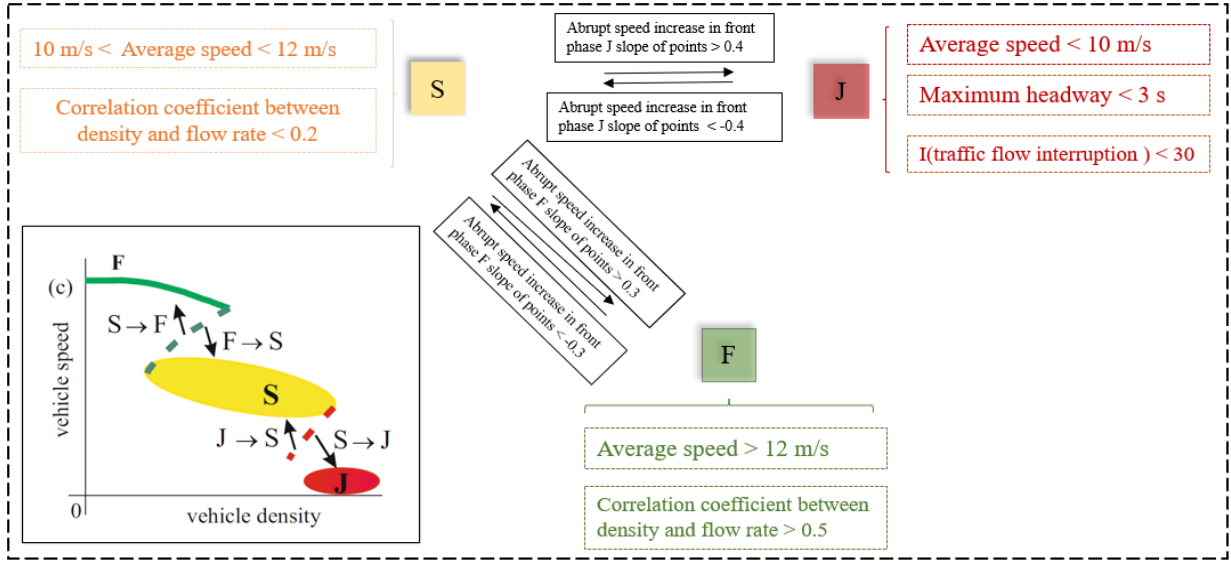

**Figure 4 Process of traffic state identification.**

The free-flow (F) phase is characterized by high speed (>12 m/s) and a strong connection between density and flow rate (> 0.5). In contrast, the correlation between density and flow rate in phase S is weak, with a correlation coefficient of less than 0.2. Meanwhile, the speed of Phase S was defined between 8 and 12 m/s. An abrupt change in speed characterizes the scenario that prevails between transitional phases F, S, and J *(33)*. There are three factors used to determine phase J. The low average speed (less than 8 m/s) was the first requirement. The maximum time headway (3s) was the second requirement. The third criterion involved a microcosmic interruption of the flow of traffic within a large moving jam: (I<30 s), where it is like the quantity of vehicles in the jam. The calculation of I is shown in **Equation 1**. Here, we set the I threshold as 30 to distinguish phase J.

$$I = \frac{\tau_J}{\tau_{jam}}, \quad (1)$$

where
$\tau_J$ is the duration of wide moving jam.
$\tau_{jam}$ is the the mean time in vehicle to pass the downstream of jam.

**Surrogate safety measures**
Traffic conflicts are used to predict interaction and the possibility of a crash if vehicles remain in their current state. Surrogate safety measures with a risk threshold can be used to assess conflicts *(34)*. It should be noted that there is no perfect conflict indicator for evaluating global conflict events. By classifying conflict indicators into three distinct types, a more profound comprehension of the interplay between conflicts and traffic flow can be attained. Rear-end collisions can be assessed using Time to Collision (TTC) as an appropriate temporal proximity indicator, providing insights into crash frequency and severity. To detect small crash probabilities and consider the road surface's friction coefficient in assessing pavement characteristics, the Proportion of Stopping Distance (PSD) serves as a valuable spatial proximity indicator. Additionally, Deceleration Rate to Avoid Collision (DRAC), which combines a vehicle's maximum available deceleration rate, proves to be a dependable kinematic indicator for predicting rear-end crash risk. As a result, conflict measures must be chosen based on the research context. This research focused on identifying conflicts and their thresholds in a simple and reliable manner without the use of a complex computation process. It should be noted that we chose the closest bonding box point between two vehicles rather than the center of the trajectory, which is more accurate to compute the SSM.

*Time-to-collision (TTC)*
Originally, Time-to-Collision (TTC) referred to the amount of time left before two vehicles would collide if they continued their current trajectory and maintained their speed difference. TTC is calculated using the **Equation 2.**

$$\text{TTC} = \frac{(X_{i-1}-X_i)-l_i}{V_i-V_{i-1}}, \tag{2}$$

where
$X$: The location of the vehicle.
$l$: The length of the vehicle.
$v$: The velocity of the vehicle.

*Proportion of Stopping Distance (PSD)*
The proportion of Stopping Distance (PSD) is defined as the quotient of the remaining distance to the potential collision point divided by the minimum acceptable stopping distance **(Equation 3).**

$$\text{PSD} = \frac{\left((P_{i-1,t}-P_{i,t})-L_{i-1}\right)}{V_i(t)^2/2\text{MADR}}, \tag{3}$$

where
$P$: The position of a vehicle.
$L$: The length of the following vehicle.
$MADR$: The maximum deceleration rate.

*Deceleration Rate to Avoid the Crash (DRAC)*
DRAC involves dividing the difference in speed between a following vehicle and a leading vehicle by the time interval between them. DRAC represents the rate at which the following vehicle needs to slow down to prevent a collision with the leading vehicle. The calculation is in **Equation 4**:

$$\text{DRAC} = \frac{(V_{i,t}-V_{i-1,t})^2}{2[(P_{i-1,t}-P_{i,t})-L_{i-1}]}, \tag{4}$$

where
$t$: Time interval.
$P$: The position of a vehicle.
$L$: The length of the following vehicle.
$v$: The velocity of the vehicle.

**Clustering methods and evaluation**
In this section, we suggest evaluating the representation of SSM through unsupervised learning using three clustering models: k-means, GMM clustering, and Mclust. Our approach involves using trajectory data as input for clustering and calculating TTC, DRAC, and PSD between vehicles. Since there are no ground-truth labels for traffic conflicts, internal evaluation methods and external evaluation methods are the two broad categories to evaluate clustering results. The external evaluation method assesses the quality of the clustering results while knowing the true label (ground truth), whereas the internal evaluation method does not rely on external information but only on the clustering results and sample attributes *(35)*. To assess the clustering outcomes, this research relies on internal metrics such as the Silhouette Coefficient, Calinski-Harabasz Score, and Davies-Bouldin Score **(Figure 5).**

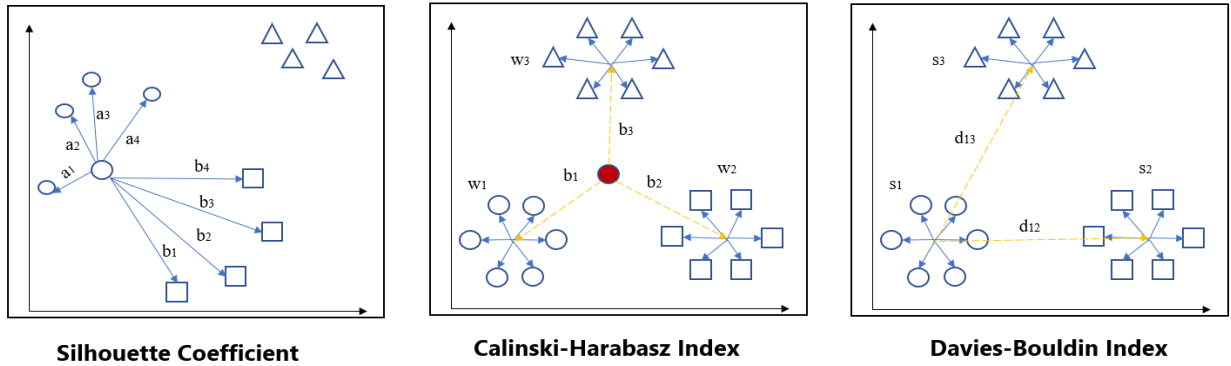

**Figure 5 Evaluation index of clustering results**

A smaller ratio of Silhouette Coefficient Index indicates a greater distance between the sample point's cluster structure and the nearest cluster structure, which implies a better clustering result *(36)*. In addition, as the Calinski-Harabasz index decreases, the distance between clusters becomes smaller, suggesting a poorer quality of clustering *(37)*. The possible values of s range from 0 to infinity. Lastly, the range of the Davies-Bouldin Index is between [0, +∞). The clustering method performs good when the index is small *(38)*.

Three models are compared from different theoretical perspectives to examine the performance of classification by unsupervised clustering. The input considers traffic conflict variables, including TTC, and DRAC. Mclust employs the Expectation-Maximization (EM) algorithm for density estimation and the Baysian Information Criterion (BIC) algorithm for model selection *(39)*. Mclust also includes model-based hierarchical clustering, which is used to initialize the EM algorithm. There are several ways to parameterize the dispersion matrix Σk. In hierarchical clustering, Fraley and Raftery presented mixture models, which assumed that maximizing the classification likelihood yields the best partitioning. Each cluster can be represented by a Gaussian model **(Equation 5).**

$$\varphi(x|\mu_i, \Sigma_i) = \frac{\exp\left(-\frac{1}{2}(x_k-\mu_i)^T \Sigma_i^{-1}(x_k-\mu_i)\right)}{\sqrt{(2\pi)^p \Sigma i}}, \qquad (5)$$

where
xk is the data points (i∈ {1, ...., m}).
k is the number of clusters.
μi and Σi reflect the mean and variance of the cluster Di where these parameters can be counted using EM algorithm with a fixed number of clusters.
Each covariance matrix is parameterized by eigenvalue decomposition **(Equation 6).**

$$\Sigma i = \lambda_i D_i A_i D_i^T, \qquad (6)$$

where
Di represents the clusters of xi.
$A_i$ is the diagonal matrix proportional to $\Sigma i$.
$\lambda_i$ is a scalar.

**RESULT**
**Description of traffic state**
The distribution of traffic flow variables at each traffic state is displayed in **Table 2**. To minimize interference, an output SSM sequence describing the interaction between each pair of vehicles was generated with a selected time step of one second (30 frames).

**TABLE 2 Description of traffic state**

| Traffic state | Num of vehicles | Duration (s) | Flow rate (veh/h/ln) | | Density (veh/m/ln) | | Speed (m/s) | | Time headway (s) | |
|---|---|---|---|---|---|---|---|---|---|---|
| | | | Mean | SD | Mean | SD | Mean | SD | Mean | SD |
| J | 287 | 267 | 342.034 | 28.524 | 1.945 | 0.133 | 4.877 | 0.798 | 2.381 | 1.845 |
| S→J | 103 | 95 | 538.751 | 147.377 | 1.814 | 0.094 | 6.979 | 0.837 | 2.069 | 0.142 |
| J→S | 144 | 75 | 554.101 | 71.177 | 1.702 | 0.166 | 7.382 | 0.663 | 2.994 | 1.32 |
| S | 406 | 1113 | 624.002 | 52.478 | 1.47 | 0.034 | 9.536 | 0.444 | 2.396 | 0.61 |
| F→S | 109 | 60 | 491.809 | 11.186 | 1.424 | 0.093 | 9.308 | 0.332 | 3.735 | 0.057 |
| S→F | 78 | 90 | 511.203 | 49.092 | 1.33 | 0.08 | 9.714 | 1.234 | 3.626 | 1.085 |
| F | 147 | 339 | 482.644 | 5.987 | 1.038 | 0.045 | 12.9 | 0.143 | 3.097 | 0.067 |

The results show that phase S has the most cars over the longest period, followed by phases J, and then F. This phenomenon occurs because there is little abrupt turbulence throughout these three steady stages. Due to traffic congestion, phase J has the lowest speed and highest density, whereas phase F has the highest speed and lowest density among them. Phase S has the maximum flow rate owing to the large number of cars during this phase. Between phases J and F, phase S has a medium speed and density. In terms of transitional states, they do not last for long, and the levels of traffic flow variables were mild, compared to stable phases (F, S, J), which include fewer vehicles. These states have higher standard deviations than other states because they are undergoing unstable transitions and turbulence of traffic flow.

**Identification of traffic conflicts and thresholds**
To validate the identification of traffic conflicts with different thresholds across various traffic states, matched traffic flow data of corresponding road segments is captured. To extract the cross-sectional traffic character data from upstream and downstream for each conflict, an upstream time is recorded from the instant entering data of every new vehicle first detected within the time interval of a conflict. If the vehicle ID cannot be founded after the ending timestamp, the end of downstream time for the corresponding vehicles will be recorded. The time of conflict is recorded by minTTC, which represents the smallest TTC value recorded among two distinct vehicle trajectories. The number of conflicts is represented by minTTC in our work. It signifies the most critical moment of potential collision between individual pairs of vehicles. Non-conflict observations are of utmost significance in studies as they showcase distinct characteristics that help discern conflict-prone situations by clustering methods. In this research, non-conflict observations refer to instances where no conflicts arise during the respective period and the subsequent timestamp (30s). The thresholds of SSMs are determined by the boundary of clusters between conflicts and non-conflicts, which is shown in **Table 3** with corresponding traffic index.

*Ding, Abdel-Aty, Zheng, Wang, and Wang***TABLE 3 Statistic summary of traffic conflicts and corresponding parameters**

| Traffic state | | all states with pre-set thresholds | all states | J | S→J | J→S | S | F→S | S→F | F |
|---|---|---|---|---|---|---|---|---|---|---|
| Threshold of TTC | | 1.500 | 3.596 | 2.901 | 3.423 | 3.500 | 3.497 | 3.628 | 3.577 | 3.998 |
| Threshold of DRAC | | 3.400 | 3.503 | 3.464 | 3.411 | 3.609 | 3.178 | 3.312 | 3.821 | 3.764 |
| Threshold of PSD | | 1.000 | 1.000 | 1.000 | 1.000 | 1.000 | 1.000 | 1.000 | 1.000 | 1.000 |
| Num of conflicts | | 173 | 322 | 87 | 31 | 23 | 42 | 34 | 18 | 21 |
| Average upstream speed (m/s) | Mean | 16.167 | 14.697 | 10.012 | 11.134 | 11.085 | 12.765 | 14.735 | 14.166 | 19.631 |
| | Standard deviation | 4.986 | 4.533 | 3.768 | 4.092 | 4.176 | 3.945 | 4.329 | 3.991 | 4.543 |
| Average downstream speed (m/s) | Mean | 16.077 | 14.615 | 9.895 | 11.352 | 12.129 | 11.478 | 13.994 | 14.263 | 19.894 |
| | Standard deviation | 5.480 | 4.982 | 2.566 | 4.864 | 3.978 | 4.189 | 5.646 | 4.564 | 5.897 |
| Difference of speed between upstream and downstream (m/s) | Mean | 3.024 | 2.749 | 2.145 | 2.356 | 2.800 | 2.012 | 2.532 | 2.128 | 3.523 |
| | Standard deviation | 2.063 | 1.875 | 1.163 | 1.843 | 1.496 | 1.238 | 1.391 | 1.472 | 3.329 |
| Standard deviation of upstream speed | Mean | 3.013 | 2.739 | 2.429 | 2.872 | 2.670 | 2.372 | 2.237 | 2.329 | 2.523 |
| | Standard deviation | 2.100 | 1.909 | 1.346 | 1.489 | 1.827 | 1.634 | 1.983 | 2.105 | 1.766 |
| Standard deviation of downstream speed | Mean | 2.937 | 2.670 | 2.297 | 2.989 | 2.526 | 2.234 | 2.303 | 2.145 | 2.496 |
| | Standard deviation | 1.337 | 1.216 | 1.223 | 1.380 | 1.842 | 1.961 | 1.721 | 1.841 | 1.236 |
| Coefficient of variation of upstream speed | Mean | 0.237 | 0.216 | 0.243 | 0.258 | 0.241 | 0.186 | 0.152 | 0.164 | 0.129 |
| | Standard deviation | 0.054 | 0.049 | 0.046 | 0.024 | 0.043 | 0.029 | 0.035 | 0.091 | 0.043 |
| Coefficient of variation of downstream speed | Mean | 0.245 | 0.222 | 0.232 | 0.263 | 0.175 | 0.208 | 0.260 | 0.150 | 0.125 |
| | Standard deviation | 0.085 | 0.077 | 0.056 | 0.097 | 0.065 | 0.086 | 0.064 | 0.043 | 0.079 |
| Upstream traffic volume (Veh/30 s) | Mean | 9.556 | 8.688 | 6.671 | 7.465 | 7.163 | 8.165 | 7.465 | 8.891 | 9.465 |
| | Standard deviation | 5.387 | 4.898 | 4.894 | 3.745 | 4.748 | 5.456 | 3.841 | 4.642 | 3.841 |
| Downstream traffic volume (Veh/30 s) | Mean | 9.684 | 8.803 | 6.784 | 7.164 | 7.984 | 8.065 | 7.135 | 9.165 | 9.723 |
| | Standard deviation | 5.630 | 5.118 | 4.215 | 4.489 | 4.921 | 4.984 | 4.413 | 4.654 | 4.895 |
| Difference of volume between upstream and downstream (Veh/30 s) | Mean | -0.127 | -0.116 | -0.113 | 0.301 | -0.821 | 0.100 | 0.330 | -0.274 | -0.258 |
| | Standard deviation | 1.281 | 1.164 | 3.256 | 0.310 | 1.416 | 0.465 | 0.894 | 0.413 | 0.654 |

Compared with all states with pre-set thresholds, using the method proposed in our work can detect more conflicts. The thresholds of TTC and DRAC are higher than the common value in previous study since it considers the traffic condition of Freeway segment. When considering the specific traffic states, we can identify more details about traffic characteristics across their transition and link the relation between traffic conflicts and traffic flow characteristics **(Figure 6).**

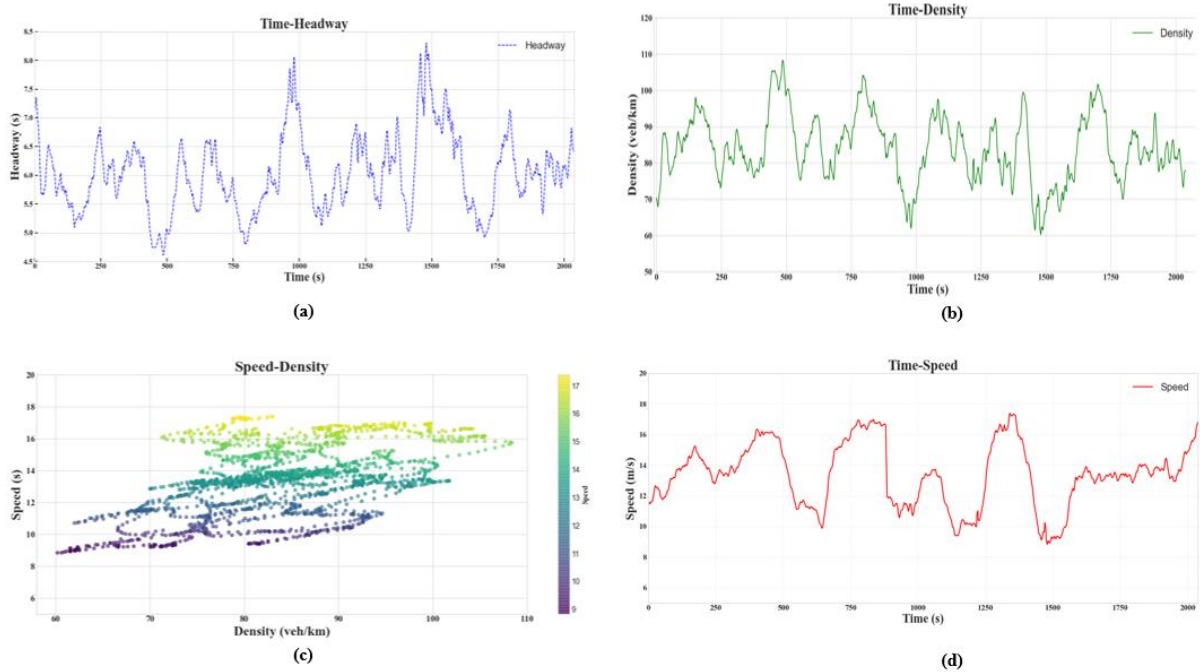

**Figure 6 Macroscopic traffic conditions (a) Time- headway plot (b) Time- density plot (c) Speed-density plot (d) Time- speed plot.**

The thresholds vary significantly among these states. Phase J has the lowest threshold of TTC, and the traffic characteristics exhibit the same tendency of distribution. The average speeds and volumes of upstream and downstream are the lowest among all the states, as well as the difference between the speed. The difference in volume between upstream and downstream is smaller than in other states, which indicates that there aren't many variations in traffic flow. The larger coefficient of variation for speed is larger during this phase, which indicates that the speed is more spread out in relation to the mean, leading to higher variability and less homogeneity. Since phase F has the greatest threshold of TTC with the same tendency of average speed and flow, which allows drivers to have more time to respond to emerging systems, making it safer than other traffic states with fewer conflicts. The smallest coefficient of variation for speed suggests that speed has less dispersion and is more homogeneous during phase F. When a vehicle is in phase S, it keeps a similar deceleration. Due to large traffic volumes in phase S, vehicles in this phase have large thresholds, which means it requires more distance for a vehicle to avoid conflict. Other than stable phases, the transition between these phases has a larger threshold compared with phase J and S. This is because the vehicle tends not to maintain the same deceleration before a crash at these phases. The turbulence of deceleration will result in changes in the remaining distance between the leading and following vehicle. The larger standard deviation and coefficient of variation of average speed suggest that the speed of different states is more spread out, and there is greater variability across the transition of traffic states. DRAC is consistent with what we find in TTC. Compared with the value ($3.5/s^2$) selected in most research, the threshold of DRAC is more accurate and sensitive to the changes in the flow and speed of vehicles. To be noticed, the threshold of PSD is selected as 1 in all the states, which is a comparison of thinking distance and braking distance, expressed as a ratio or percentage. Based on the PSD, TTC, and DRAC values obtained from the Citysim dataset, phase J poses the highest risk of conflict when traffic flow is extremely heavy and congested. Phase S follows with the second-highest conflict risk, which may be due to the high density and small space headway between surrounding vehicles. The maximum space headway between vehicles explains why phase F has fewer conflicts than S and J, as shown in Figure 6. Moreover, due to the various traffic features, the transitional stages between these phases experience more conflicts than phase F. Hazardous situations may arise during the transitional state, as drivers tend to alter their behavior by decelerating in response to stop-and-go waves, which can exacerbate conflicts. The risk of conflict is higher during the S→J and F→S transitional states than in other transitional states. The high flow rate and vehicle speed during the F→S transitional state make it significantly more dangerous than phase F.

**CONCLUSIONS**

In this study, the three-phase traffic theory is utilized to establish a link between macroscopic traffic flow states and microscopic traffic conflicts. By analyzing microscopic traffic trajectory data, an unsupervised clustering method is proposed in this research to detect traffic conflicts and establish the SSM thresholds based on the three-phase framework. Initially, traffic states and their transitions are identified using the three-phase theory and traffic characteristics. Conflicts in each state of traffic were then assessed using SSMs including TTC, DRAC, and PSD. After comparing various clustering methods, the conflicts and thresholds were clustered using Mclust method.

High-resolution trajectory data was applied to validate the method. The strategy suggested in this research keeps most of the current occurrences and is not influenced by external parameters, in contrast to the correlational approach. Compared with extreme value estimation, the proposed method doesn't require a long-term data and can be calculated without crash observations. The study indicates that phase J poses the highest risk of conflicts based on the conflict outcomes, with phase S following closely due to its substantial sample size. On the other hand, phase F demonstrates better performance than the other phases. The transitional states exhibit comparable levels of conflict risk, with the S→J and F→S transitions displaying more conflicts than the other transitions. These results suggest that the distribution of traffic conflict varies depending on the traffic state. Meanwhile, the thresholds of traffic conflicts cannot be fully represented by the same value through different traffic states.

Furthermore, rather than using a single threshold as a specific measure, this method detected conflicts by analyzing SSM sequences from multiple perspectives, demonstrating the advantage of diverse thresholds over one. Compared with other approaches to identify thresholds, we can classify conflicts and non-conflicts in an unsupervised but more reliable manner. Another contribution of this study is to compute SSMs more accurately with closest points of the bounding box instead of the centers of the vehicle trajectory. The sample of empirical traffic data used in our research is limited, comprising only 35 minutes of recordings. To gain more robust insights into various traffic scenarios like signalized intersections, ramp vicinities, and roundabouts, it is crucial to incorporate trajectory data with a larger sample size and greater diversity. Additionally, future research endeavors could focus on refining the conflict prediction model by considering factors such as weather conditions, geometric designs, and other variables to address potential issues of endogeneity in conflict analysis.

**AUTHOR CONTRIBUTIONS**
The authors confirm contribution to the paper as follows: study conception and design: Shengxuan Ding, Mohamed Abdel-Aty, Ou Zheng, Zijin Wang, Dongdong Wang; data collection and processing: Shengxuan Ding, Ou Zheng; analysis and interpretation of results: Shengxuan Ding, Zijin Wang, Dongdong Wang; draft manuscript preparation Shengxuan Ding, Mohamed Abdel-Aty, Ou Zheng, Zijin Wang, Dongdong Wang. All authors reviewed the results and approved the final version of the manuscript.